\documentclass[reprint,
groupedaddress,
 twocolumn,
 nofootinbib,
 nobibnotes,
 longbibliography,
 amsmath,amssymb,
 aps,
pra,
 floatfix,
 tightenlines,
 10pt
]{revtex4-1}

\usepackage{physics}
\usepackage{xcolor}
\usepackage{float}
\usepackage{graphicx}
\usepackage{subcaption}
\usepackage{ulem}
\usepackage{soul}
\usepackage{caption}
\usepackage[colorlinks=true,citecolor=blue,urlcolor=blue,linkcolor=blue]{hyperref}
\usepackage{relsize}
\usepackage{xcolor}
\usepackage{amsmath, nccmath}

\RequirePackage{stix}

\newcommand{\figref}{FIG. \ref}

\newcommand{\appendixref}{Appendix \ref}
\renewcommand{\eqref}[1]{{Eq.~(\ref{#1})}}

\newcommand{\SC}{{\rm{S}}}
\newcommand{\normal}{{\rm{N}}}
\newcommand{\hc}{{\rm{H.c.}}}
\newcommand{\inter}{{\rm{int}}}
\newcommand{\B}{{\rm{B}}}

\usepackage{bm}

\begin{document}
\title{Elevated critical temperature at BCS superconductor--band insulator interfaces}

\author{Mats Barkman}
\affiliation{Department of Physics, Royal Institute of Technology, SE-106 91 Stockholm, Sweden}

\author{Albert Samoilenka}
\affiliation{Department of Physics, Royal Institute of Technology, SE-106 91 Stockholm, Sweden}

\author{Andrea Benfenati}
\affiliation{Department of Physics, Royal Institute of Technology, SE-106 91 Stockholm, Sweden}

\author{Egor Babaev}
\affiliation{Department of Physics, Royal Institute of Technology, SE-106 91 Stockholm, Sweden}

\begin{abstract}

We consider the interface between a Bardeen-Cooper-Schrieffer
superconductor and non-superconducting band insulator. 
We show that under certain conditions, such interfaces can have an
elevated superconducting critical temperature, without increasing the strength of the pairing interaction at the interface.
We identify the regimes where the interface critical temperature   exceeds the critical temperature associated with a superconductor-vacuum interface.
\end{abstract}

\maketitle

\section{Introduction}

A series of classical works by de Gennes \textit{et. al} \cite{book_de_Gennes,deGennes_Boundary,CdGM_Coherence,CdGM_french} considered the interface between a Bardeen-Cooper-Schrieffer (BCS) superconductor and a normal material without electron pairing.
The calculations predicted superconducting gap suppression at the interface.
The same calculation predicted that in the limit where the material of the interface is dielectric, the normal derivative of the superconducting gap becomes zero. Hence in that limit, superconductivity is neither enhanced nor suppressed near such an interface (similar  results were obatined in \cite{abrikosov1965concerning,andryushin1993boundary}).
The zero normal derivative of the superconducting gap is widely considered to be independent of the specifics of the dielectric involving $s$-wave superconductors unless
  the superconductor is anisotropic  (for a review see \cite{andryushin1993boundary}). In the
  anisotropic case the superconducting gap can be suppressed near the dielectric interface \cite{shapoval1985boundary,andryushin1993boundary}.

The problem of the boundary between a superconductor and vacuum was recently revisited. It was shown  
 in  \cite{samoilenka2020boundary,samoilenka2020pair,samoilenka2021microscopic} that  there are boundary states in a BCS superconductor that have  higher critical temperature than {the critical temperature of the bulk}. A rigorous mathematical proof of that result was recently presented in \cite{hainzl2022boundary}. These are highly inhomogeneous   solutions for the superconducting order parameter $\Delta ({\bf r})$, localized on a macroscopic lengths scale near the surface. 
 The origin of the effect is rooted in interference effects (that were neglected in \cite{book_de_Gennes,deGennes_Boundary,CdGM_Coherence,CdGM_french,abrikosov1965concerning}) arising when electrons scatter from a perfectly reflecting surface.
The effect requires the solution of the full microscopic model and is not captured in a straightforward application of quasiclassical approximation or Ginzburg-Landau model.
However, effective models can capture these states with appropriate microscopically derived boundary conditions \cite{samoilenka2021microscopic}.
There are various degrees of experimental evidence of enhanced surface superconductivity  in various superconductors \cite{fink1969surface,lortz2006origin, janod1993split,khlyustikov2011critical,khlyustikov2016surface,kozhevnikov2007observation,khlyustikov2021surface,mangel2020stiffnessometer,tsindlekht2004tunneling,belogolovskii2010zirconium,khasanov2005anomalous}.

In this work we revisit the problem of the interface between a superconductor and normal material. We show that interfaces between BCS superconductors and band insulators, depending on the nature of the dielectric, can have elevated critical temperatures, without the introduction of a new boundary pairing mediator (the effect of additional pairing mediators at interfaces was considered in \cite{ginzburg1964surface}).
The critical temperature of such an interface can even exceed the critical temperature of the perfectly reflective superconductor-vacuum boundary
 \cite{samoilenka2020boundary,samoilenka2020pair,samoilenka2021microscopic}.

\section{Model}

\begin{figure}
    \centering
    \includegraphics[width=\columnwidth]{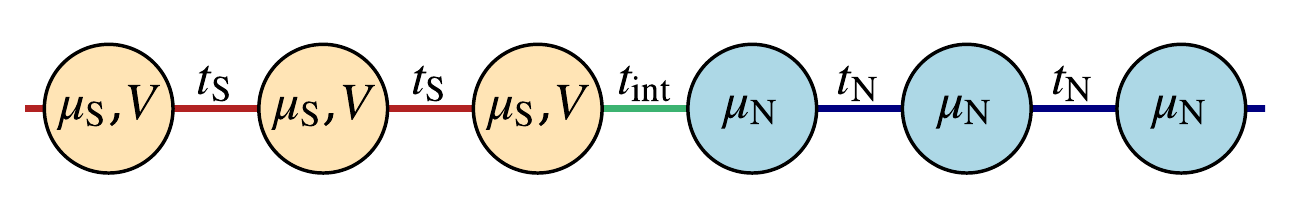}
    \caption{Illustration of the lattice model for the superconductor-normal interface. The superconductor (to the left) has nonzero pairing potential $V$, on-site potential $\mu_\SC$, and hopping parameter $t_\SC$. Similarly the normal material (to the right) has some on-site potential $\mu_\normal$ and hopping parameter $t_\normal$, but no pairing interaction. The two materials are linked through a hopping parameter $t_\inter$.
    }
    \label{fig: model sketch}
\end{figure}

We model the superconducting-normal interfaces using a mean-field microscopic lattice model, described by the Hamiltonian
\begin{equation} \label{eq: mean-field hamiltonian}
    \begin{aligned}
    H = &   - \sum_{ \sigma, <x x'>} t(x,x')  c^\dagger_{ \sigma } (x) c_{\sigma } (x') -\sum_{\sigma, x} \mu (x) c^\dagger _{\sigma} (x) c_{\sigma} (x)   \\
    & +\sum_{x} \left( \Delta (x) c^\dagger_{\uparrow} (x) c^\dagger_{\downarrow} (x) + \hc \right),
    \end{aligned}
\end{equation}
where $c_{\sigma} (x)$ is the annihilation operator of an electron with spin $\sigma$ at position $x$, $<xx'>$ denotes nearest-neighbor pairs, and $\hc$ denotes Hermitian conjugation. The hopping parameters $t (x,x')$ and the on-site chemical potentials $\mu (x)$ are not necessarily the same in both materials. The hopping parameter can take three values: $t_\SC$ in the superconductor, $t_\inter$ at the interface, and $t_\normal$ in the nonsuperconducting material. Analogously, the on-site chemical potential can either be equal to $\mu_\SC$ or $\mu_\normal$. This is illustrated in \figref{fig: model sketch}.
The superconducting pairing amplitude $\Delta (x)$ is defined through the thermal average
\begin{equation} \label{eq: definition delta}
    \Delta (x) = V (x) \expval{c_{\uparrow} (x) c_{\downarrow}(x)},
\end{equation}
where $V (x)$ is the pairing potential which equals zero in the normal material.
In this paper we are interested in computing the critical temperature of the interface. 
For a {phase transition where} $\Delta (x)$ vanishes {continuously} at criticality, the standard approach is to use a linearized gap equation to find the critical temperature. The linearized gap equation reads
\begin{equation} \label{eq: linearised gap equation}
    \Delta (x) = V(x) \sum_{x'} K(x,x') \Delta (x'), 
\end{equation}
where $K(x,x')$ can be expressed in terms of the wavefunctions $\phi_{\sigma n} (x)$ in the absence of superconductivity and their corresponding eigenvalues $\epsilon_{\sigma n}$ as
\begin{equation} \label{eq: K matrix definition}
    K(x,x') = \sum_{m,n} F_{mn}\phi_{\uparrow m} (x) \phi_{\downarrow n} (x) \phi_{\uparrow m}^* (x') \phi_{\downarrow n}^* (x'),
\end{equation}
and
\begin{equation} \label{eq: F matrix definition}
    F_{mn} = \frac{1-f(\beta\epsilon_{\uparrow m}) - f(\beta\epsilon_{\downarrow n})}{\epsilon_{\uparrow m} + \epsilon_{\downarrow n}},
\end{equation}
where $f(z) = [1+\exp(z)]^{-1}$ is the Fermi-Dirac distribution function and $\beta = (k_{\rm{B}}T)^{-1}$ is the inverse temperature. At the critical temperature $T_c$, the largest eigenvalue to the matrix $V(x) K(x,x')$ equals 1. {We determine the wavefunctions and their corresponding eigenvalues numerically, from which the matrix $K(x,x')$ can be constructed and the critical temperature computed.}

\section{Results}
\begin{figure}
    \centering
    \includegraphics[width=\columnwidth]{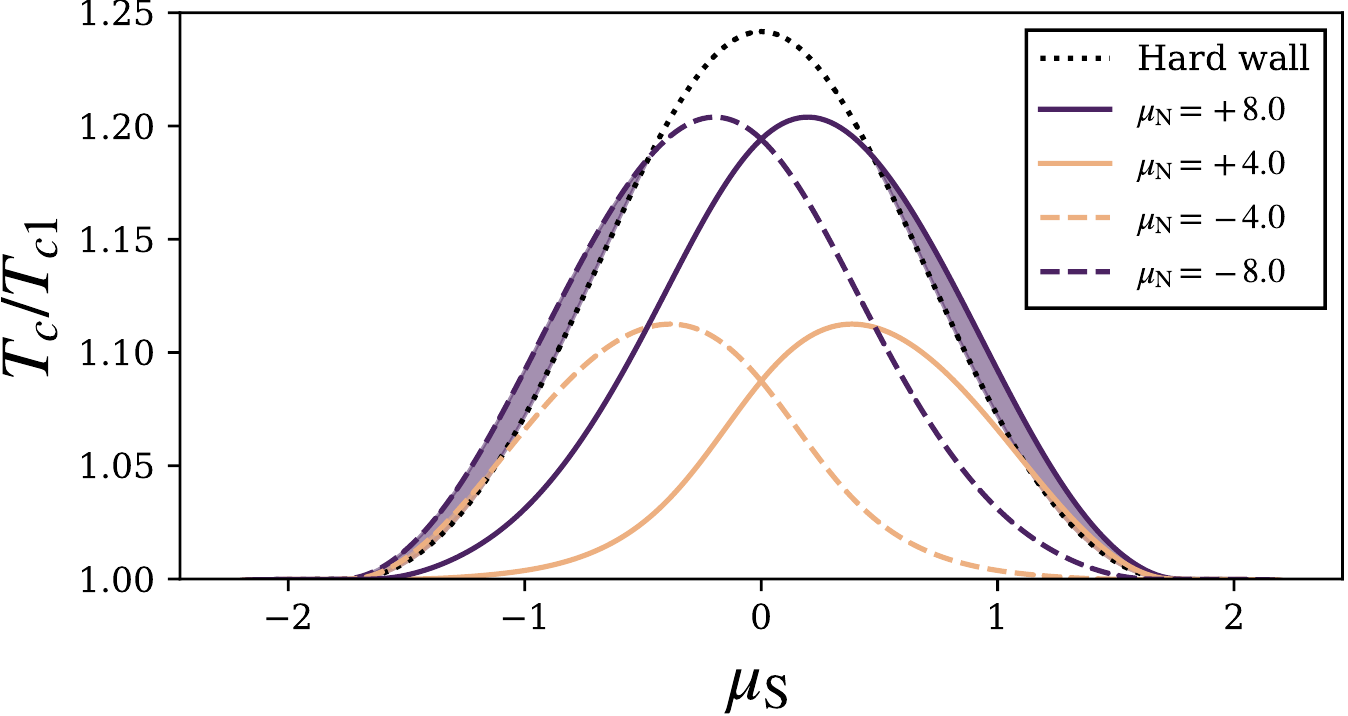}
    \caption{Interface critical temperature $T_c$ in units of the bulk critical temperature $T_{c1}$ for some fixed on-site potentials $\mu_\normal$ in the normal material, as a function of the on-site potential $\mu_\SC$ in the superconductor. The dotted line corresponds to the hard-wall boundary critical temperature $T_{c2}$. It shows that    it is possible that the interface between a superconductor and band insulator can have higher critical temperature than a superconductor-vacuum interface. All hopping parameters $t_\SC = t_\normal = t_\inter = 1$ and superconducting pairing potential $V = 2$. }
    \label{fig: Tc / Tc1 same hoppings}
\end{figure}

\begin{figure}
    \centering
    \includegraphics[width=\columnwidth]{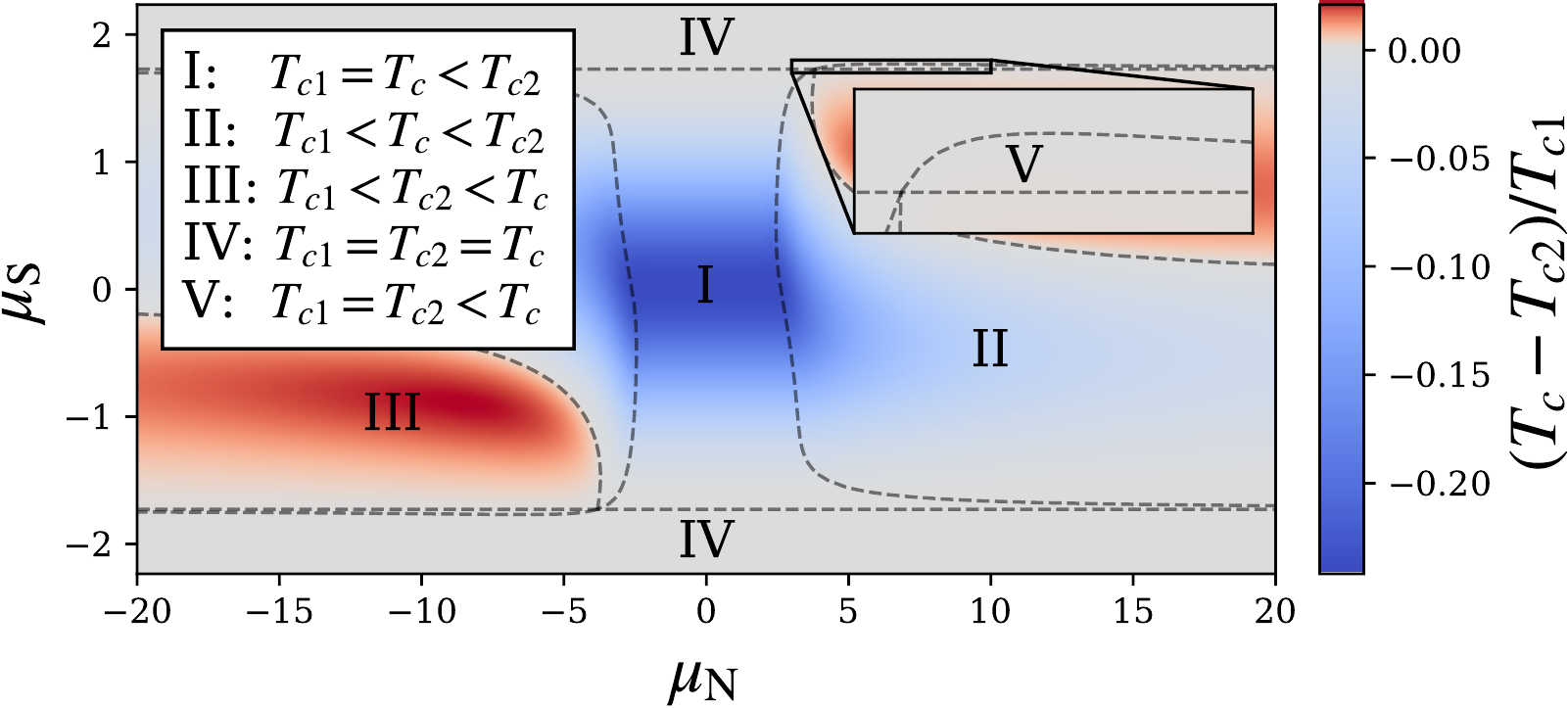}
    \caption{
    Difference between the superconductor--normal interface critical temperature $T_c$ and the critical temperature of  superconductor-vacuum interface $T_{c2}$ (in units of bulk critical temperature $T_{c1}$) for various on-site potentials $\mu_\SC$ and $\mu_\normal$ in the superconductor and the normal material respectively.
    The blue (red) parameter regime shows where the interface critical temperature is lower (higher) than the critical temperature of a superconductor-vaccuum boundary. The plane is partitioned into five regions, based on the relation between the interface,  hard-wall, and bulk critical temperatures ($T_c$, $T_{c2}$, and $T_{c1}$, respectively). All hopping parameters $t_\SC = t_\normal = t_\inter = 1$ and superconducting pairing potential $V = 2$.} \label{fig: regimes larger smaller than Tc2}
\end{figure}

Let us study the influence on the interface critical temperature from the on-site potential $\mu_\normal$ in the nonsuperconducting material. We measure all energies in units of the hopping parameter $t_\SC$ in the superconductor, and therefore set $t_\SC = 1$. For simplicity, let us assume that all other hopping parameters $t_\inter=t_\normal=1$. This leaves us with three variables: the pairing potential $V$ and the two on-site potentials $\mu_\SC$ and $\mu_\normal$. We  consider coupling strength $V=2$, for which bulk superconductivity  is present for $|\mu_\SC| \lesssim 2.2357$. For values outside this range, both the bulk critical temperature $T_{c1}$ and the hard-wall boundary critical temperature $T_{c2}$ {(corresponding to $t_\inter=0$)} vanish. Note also that the boundary critical temperature $T_{c2}$ is larger than the bulk critical temperature $T_{c1}$ only for $|\mu_\SC| \lesssim 1.7284$. We compute the critical temperature of the superconductor-normal interface for various values of the on-site potential $\mu_\normal$ in the normal material. The resulting critical temperatures are shown in \figref{fig: Tc / Tc1 same hoppings}. Note that $T_c (\mu_\SC, \mu_\normal) = T_c (-\mu_\SC, -\mu_\normal)$, which is a consequence of particle-hole symmetry. At half-filling in the superconductor, that is $\mu_\SC = 0$ $\left(\expval{n_\uparrow + n_\downarrow} = 1\right)$, increasing the magnitude of the on-site potential has the effect of increasing the interface critical temperature $T_c$ from $T_{c1}$ at $\mu_\normal = 0$, to $T_{c2}$ as $\mu_\normal \to \infty$. For nonzero on-site potential in the superconductor, beyond half-filling, the situation is different. We notice that the interface critical temperature can exceed the hard-wall critical temperature $T_{c2}$ for some values of the on-site potential $\mu_\normal$. For positive values of $\mu_\SC$, this occurs when $\mu_\normal$ also is positive. 
We show for which parameters the interface critical temperature exceeds the hard-wall critical temperature in \figref{fig: regimes larger smaller than Tc2}. 
We notice five different parameter regimes. For low magnitudes of the normal on-site potential $\mu_\normal$, the interface critical temperature does not exceed the bulk critical temperature (region I). As the magnitude of  $\mu_\normal$ increases, so does the interface critical temperature. The interface critical temperature exceeds the bulk critical temperature in region II and increases beyond the hard-wall critical temperature in region III. As the magnitude of $\mu_\normal$ approaches infinity, the interface critical approaches the hard-wall critical temperature from above in region III and from below in region II. Region IV corresponds to the parameters where all the three critical temperatures are equal. 
{There also exists a fifth region where the hard-wall and bulk critical temperatures are equal and slightly smaller than the interface critical temperature. This regime is very narrow for this particular case.} To summarize, we find that the interface between a BCS superconductor and a normal material can have a  critical temperature that  exceeds the bulk critical temperature, even without additional pairing mediators at the interface. For the considered model, it is necessary that $|\mu_\normal| \gtrsim 2.45$, which equates to the condition that the non-superconducting material is  insulating {(since $|\mu / t|>2$ leads to a completely filled or empty band at zero temperature)}. The critical temperature of the interface can furthermore be larger than the perfectly reflective superconductor-vacuum boundary, for larger values of $|\mu_\normal|$.
{In \mbox{\appendixref{appendix: example 2d}} and \mbox{\figref{fig: example enhanced Tc in 2d}} we demonstrate enhanced interface critical temperatures for two-dimensional materials. This indicates that the effects shown in the one-dimensional case also are present in higher dimensions.}

\begin{figure}
    \centering
    \includegraphics[width=\columnwidth]{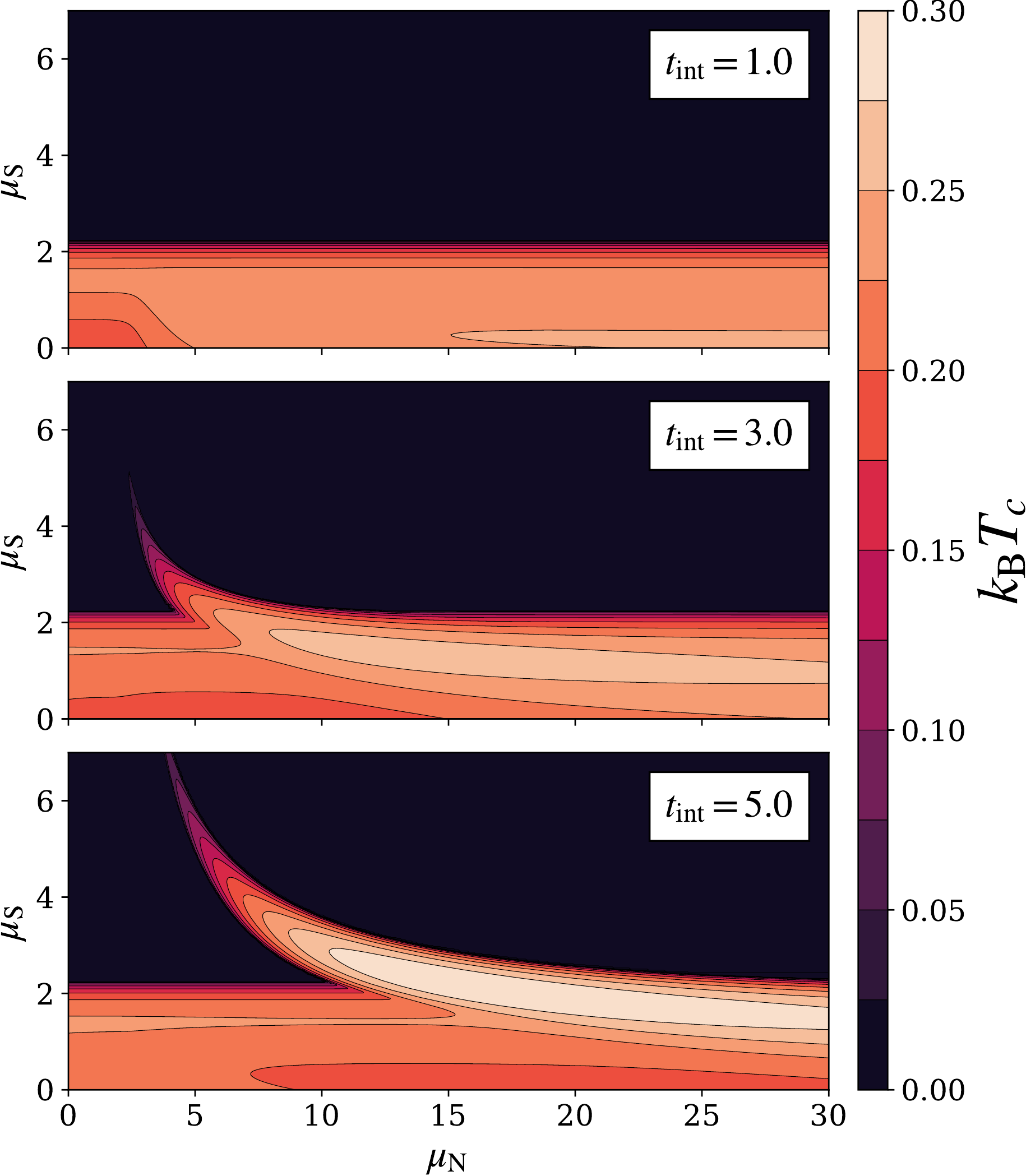}
    \caption{Interface critical temperature for different values of the interface hopping parameter $t_\inter$. As $t_\inter$ increases, interface superconductivity can arise in parameter regimes which do not support bulk superconductivity (for $|\mu_\SC|\gtrsim 2.2357$). Increasing the interface hopping parameter results in larger critical temperatures and wider parameter regimes that support interface superconductivity. Remaining parameters are set to $V=2$ and $t_\SC = t_\normal = 1$.} \label{fig: interface Tc different Tc contour}
\end{figure}

In the previous example, we studied an interface where we only changed the on-site potentials, keeping all hopping parameters equal.  However, the hopping parameter at the interface is, in general, different from the hopping parameters inside the materials.  To consider a more general situation, we will now vary the hopping parameter $t_\inter$ at the interface, along with the on-site potentials, keeping the hopping parameters in the superconductor and the normal material $t_\SC = t_\normal = 1$ fixed. Let us restrict ourselves to positive on-site potentials.
In the case when all hopping parameters are equal, as in \mbox{\figref{fig: regimes larger smaller than Tc2}}, we do observe enhanced interface critical temperatures, however, only for on-site potentials $\mu_\SC$ in the superconductor for which there is also superconductivity in the bulk (for $|\mu_\SC|\lesssim 2.2357$). Outside this parameter range, the bulk and hard-wall boundary critical temperature equal zero. However, if the interface hopping parameter $t_\inter$ differs from the hopping parameter in the superconductor, the interface critical temperature can be nonzero even when the bulk critical temperature is zero. This is shown in \mbox{\figref{fig: interface Tc different Tc contour}}, where the interface critical temperature is nonzero for larger values of $\mu_\SC$. For increasing values of $\mu_\SC$, the critical temperature decreases and the regime in $\mu_\normal$ with nonzero critical temperatures becomes more narrow. Increasing the interface hopping parameter $t_\inter$ results in enhanced critical temperatures and larger parameter regimes that support interface superconductivity.

Having demonstrated that superconductor-band insulator interfaces support enhanced critical temperatures, the natural next question is by how much the critical temperature of such interfaces can be enhanced (without additional interface pairing mediators).
Let us begin with keeping both the hopping parameters in the superconducting and normal materials fixed at $t_\SC = t_\normal = 1$. 
For each on-site potential $\mu_\SC$ in the superconductor, we want to find the normal on-site potential $\mu_\normal$ and the interface hopping $t_\inter$ such that the interface critical temperature is maximal.
The maximal interface critical temperature is obtained in the limit where both $t_\inter$ and $\mu_\normal$ are large, while keeping the ratio $t_\inter ^2 / \mu_\normal$ constant (for details see \mbox{\appendixref{appendix: extrapolation}}).
We can also check that the limiting critical temperature and optimal ratio $t_\inter^2 / \mu_\normal$ does in fact not depend on the hopping parameter $t_\normal$ in the normal material. This holds for all values of the superconducting on-site potential $\mu_\SC$, which allows us to compute the maximal interface critical temperature for each $\mu_\SC$ (the optimal ratio $t_\inter^2 / \mu_\normal$ does of course also depend on $\mu_\SC$).

\begin{figure}[t]
    \centering
    \includegraphics[width=\columnwidth]{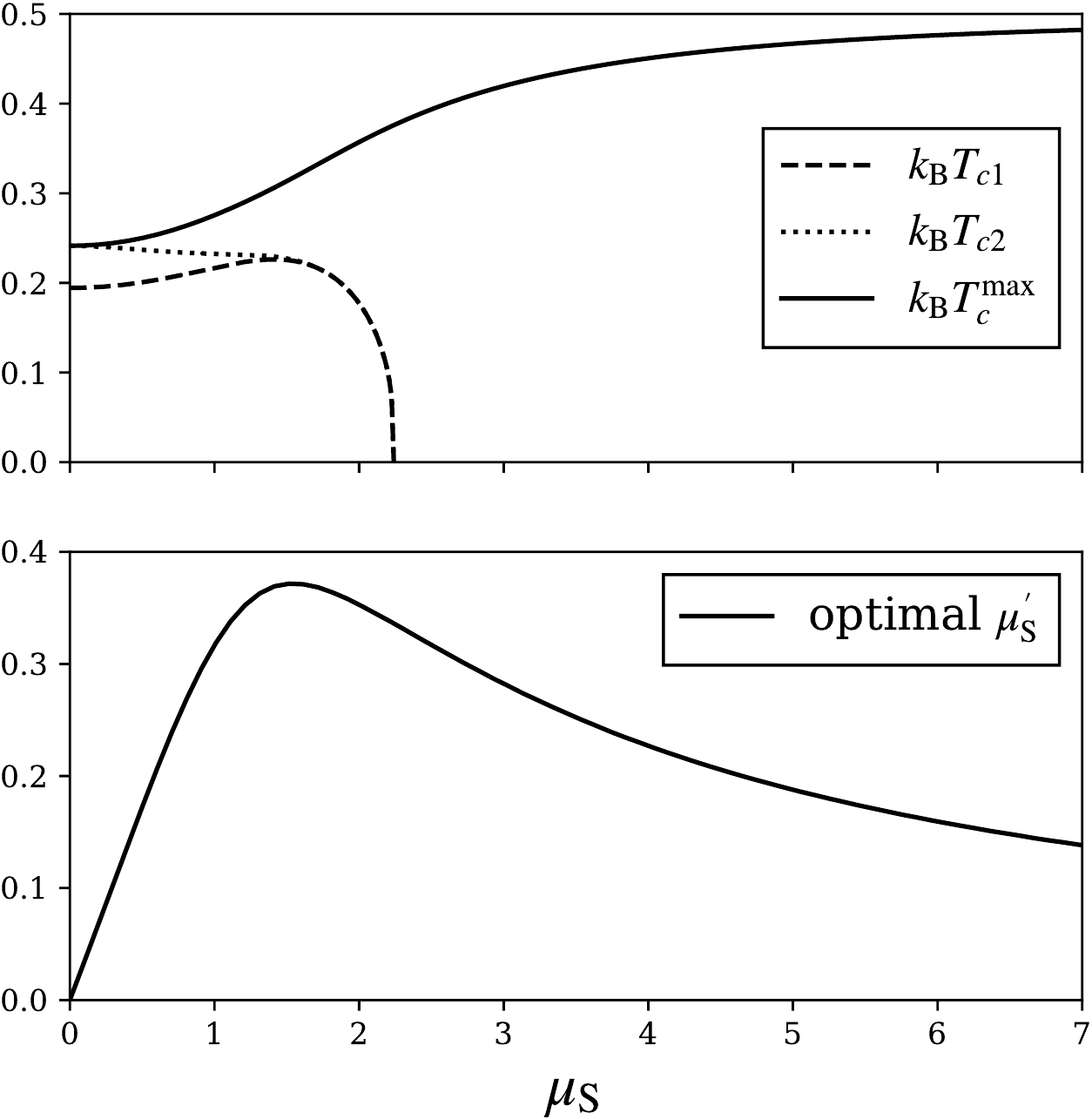}
    \caption{{Upper panel:} Bulk critical temperature $T_{c1}$, hard-wall critical temperature $T_{c2}$ and maximal interface critical temperature $T_c^{\rm{max}}$ for different values of the on-site potential $\mu_\SC$ in the superconductor. Both the bulk and hard-wall critical temperature approach zero simultaneously, while the maximal interface critical temperature increases, approaching its asymptote $V/4$ as $\mu_\SC$ approaches infinity.
    {Lower panel: Optimal shifted on-site chemical potential $\mu_\SC '$at the interface boundary. As $\mu_\SC$ approaches infinity, $\mu_\normal '$ goes to zero.} The superconducting pairing potential $V=2$ and the hopping parameter $t_\SC = 1$.}
    \label{fig: Tc1 Tc2 TcMax}
\end{figure}

We have shown that to maximize the  critical temperature of the interface with a normal material, one should take the limit of large $\mu_\normal$ and $t_\inter$ at a constant ratio $t_\inter  ^2 / \mu_\normal$.
The optimal value of this ratio depends on the on-site potential in the superconductor $\mu_\SC$, but is independent of the hopping parameter $t_\normal$ in the normal material. 
The origin of this result can be understood by analyzing the linear gap equation. The critical temperature is uniquely determined by the superconducting pairing potential $V(x)$ and the wavefunctions $\phi_{\sigma n}(x)$ and their eigenvalues $\epsilon_{\sigma n}$ in the absence of superconductivity. In the limit  where both $t_\inter$ and $\mu_\normal$ are large, only the states that are predominantly localized inside the superconductor give non-negligible contributions to the linear gap equation. These wavefunctions decay exponentially inside the normal material on a extremely short lengthscale. Therefore it is sufficient to consider only one site of normal material. Let $x=0$ be the last superconducting site and $x=1$ first and only relevant normal site. The Schr{\"o}dinger equation at site $x=1$ reads
\begin{equation}
    -t_\inter \phi (0) - \mu_\normal \phi (1) -t_\normal \phi(2) = \epsilon \phi (1),
\end{equation}
{where we dropped the indices $\sigma$ and $n$ for brevity. As stated above, $\phi (2)$ is small and can be neglected. By expressing $\phi (1)$ in terms of $\phi (0)$ and inserting this relation into the equation at site $x=0$ gives}
\begin{equation} \label{eq: effective hard-wall equation}
   - t_\SC \phi(-1) -\left(\mu_\SC - \frac{t_\inter ^2}{\mu_\normal + \epsilon} \right) \phi (0) = \epsilon \phi (0).
\end{equation}
{The ratio $t_\inter ^2 /(\mu_\normal + \epsilon) \simeq t_\inter^2 / \mu_\normal$ in the limit of large $\mu_\normal$. We can identify {\eqref{eq: effective hard-wall equation}} as the equation for hard-wall interface (i.e., superconductor-vacuum boundary), where the on-site potential on the boundary site has been shifted. This shows that, in this particular limit}, solving the linear gap equation for the full superconductor-normal interface is equivalent to solving the linear gap equation for a superconductor-vacuum boundary, where the on-site potential on the last superconducting site was shifted to $\mu_\SC ' = \mu_\SC - t_\inter  ^2 / \mu_\normal$. 
We can use this equivalence to easily compute the optimal ratio $t_\inter  ^2 / \mu_\normal$ (or equivalently optimal $\mu_\SC '$) and the maximal interface critical temperature, for each on-site potential $\mu_\SC$ in the superconductor. 
{This shift in the on-site potential at the boundary changes locally the density of states and will affect the interface critical temperature.} 
The resulting interface critical temperature is shown in \figref{fig: Tc1 Tc2 TcMax}, along with the bulk and hard-wall critical temperatures $T_{c1}$ and $T_{c2}$. At half-filling ($\mu_\SC = 0$), the maximal interface critical temperature equals the hard-wall critical temperature.
As $\mu_\SC$ increases, note that both the bulk and hard-wall critical temperature approaches zero simultaneously, while the interface between these non-superconducting materials remains superconducting and  the corresponding critical temperature increases with  $\mu_\SC$. 
{In the limit of large $\mu_\SC$}, it is sufficient to study superconductivity only at the last superconducting site, where the effective on-site potential was shifted. The linear gap equation for this single site becomes
\begin{equation} \label{eq: linear gap equation one site}
    1 = V \frac{\tanh \left( \frac{\mu_\SC '}{2 k_\B T_c} \right) }{2 \mu_\SC '},
\end{equation}
where $\mu_\SC ' = 0$ results in the maximal critical temperature $k_\B T_c = V / 4$. We can indeed see that this value is approached as $\mu_\SC \to \infty$ in \figref{fig: Tc1 Tc2 TcMax} (where $V=2$).

\begin{figure}[t]
    \centering
    \includegraphics[width=\columnwidth]{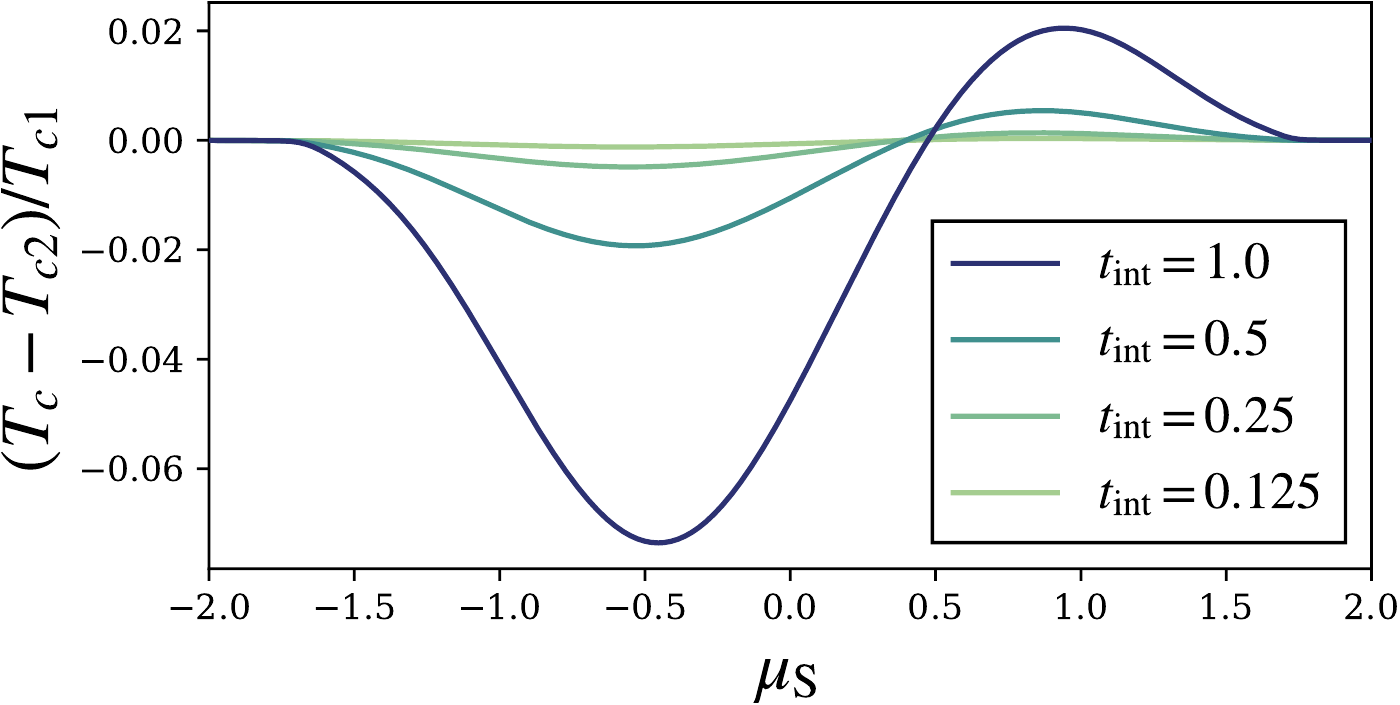}
    \caption{{Difference between interface critical temperature $T_{c}$ and hard-wall critical temperature $T_{c2}$ (in units of bulk critical temperature $T_{c1}$), for weakly coupled interfaces (small $t_\inter$), and for a fixed on-site potential $\mu_\normal=8$ in the non-superconducting material. This shows that, even for weakly coupled interfaces, there is still a difference between the hard-wall critical temperature and the interface critical temperature, although less pronounced.  The hopping parameters $t_\SC = t_\normal = 1$ and the superconducting pairing potential $V=2$.}} \label{fig: weak interface}
\end{figure}

{We showed that, to additionally enhance  the interface critical temperature, the interface hopping parameter $t_\inter$ should be large. For completeness case, let us also study the limit of small $t_\inter$, that is weakly coupled superconductor-normal interfaces. We know that when $t_\inter = 0$, the interface critical temperature equals the hard-wall critical temperature $T_{c2}$. We show the difference between the interface and hard-wall critical temperatures in \mbox{\figref{fig: weak interface}} for $t_\inter \leq 1$, and for $\mu_\normal=8$. We see that even for small $t_\inter$, the two critical temperatures differ, and the interface critical temperature can exceed the hard-wall critical temperature. As expected, the difference approaches zero as $t_\inter \to 0$.}

\section{Conclusion}

In conclusion, we revisited the problem of a boundary between a BCS superconductor and a non-superconducting material.
We showed that when the non-superconducting material is a band insulator
 the interface can acquire 
an elevated  superconducting critical temperature.
The effect arises in basic BCS theory (i.e., without the introduction of a new interface pairing mediator) and is closely connected with the nature of electronic scattering from the interface.
The critical temperature of a superconductor-band insulator interface is in general different from, and can exceed, the elevated critical temperature associated with a perfectly reflective superconductor-vacuum boundary \cite{samoilenka2020boundary,samoilenka2020pair,samoilenka2021microscopic}.
This suggests investigating granular materials with well-insulating oxides and superconductor-insulator metamaterials as a possible route to engineer improved superconducting properties.

\section{Acknowledgments}
The work was supported by the Swedish Research Council Grants  2016-06122, 2018-03659.

\begin{figure}[t]
    \centering
    \includegraphics[width=\columnwidth]{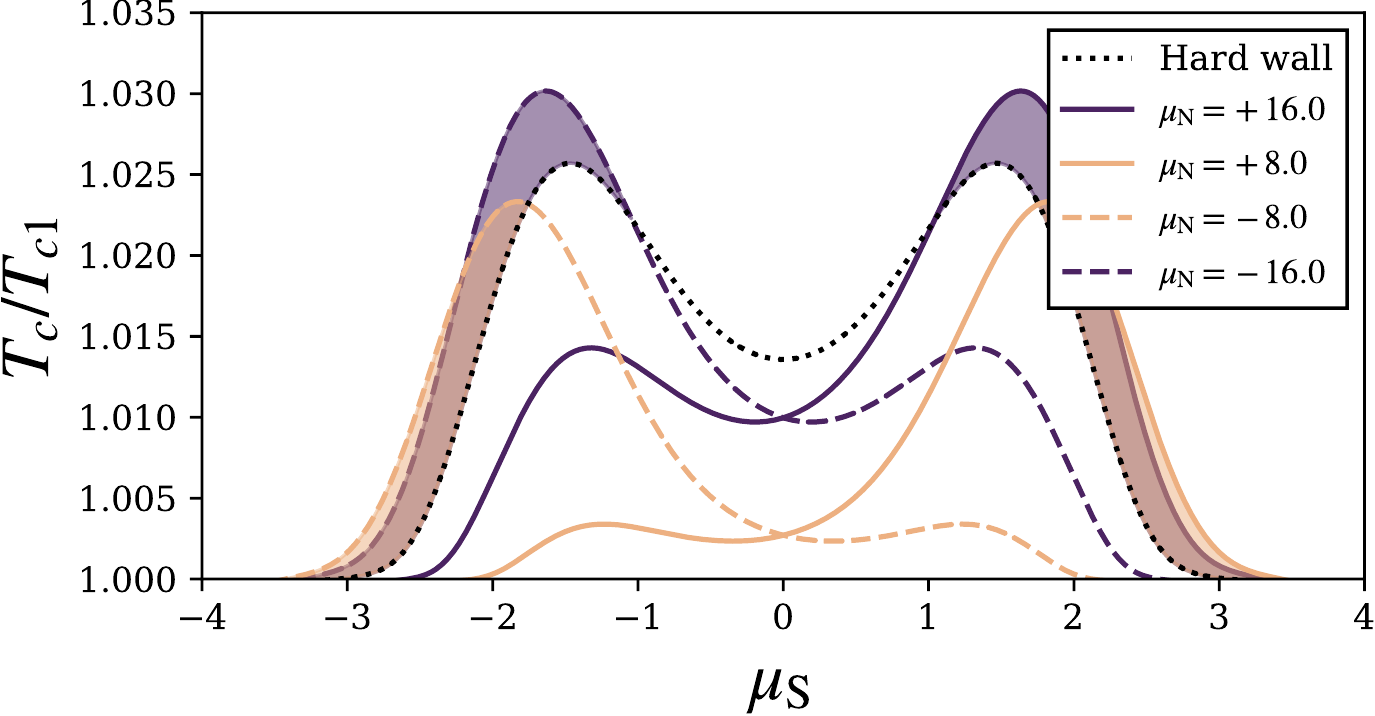}
    \caption{{Example of enhanced interface critical temperature $T_c$ (in units of bulk critical temperature $T_{c1}$) for a two-dimensional square lattice. Similarly as in the one-dimensional case in \mbox{\figref{fig: Tc / Tc1 same hoppings}}, beyond half-filling there exists regimes where the interface critical temperature can exceed the hard-wall critical temperature. The hopping parameters $t_\SC = t_\normal = t_\inter = 1$ and the superconducting pairing potential $V=3$.}} \label{fig: example enhanced Tc in 2d}
\end{figure}

\appendix
\section{Example in the two-dimensional case} \label{appendix: example 2d}

{Here we show that the enhancement of interface critical temperature does not only occur in the simplest one-dimensional case (studied in the main text), but is also present in the two-dimensional case. Similarly as in \mbox{\figref{fig: Tc / Tc1 same hoppings}}, we study the interface critical temperature for the two-dimensional interface in \mbox{\figref{fig: example enhanced Tc in 2d}} for different chemical potentials in the superconductor and the normal material. We find that the interface critical temperature can exceed both the bulk critical temperature $T_{c1}$ and the hard-wall critical temperature $T_{c2}$.}

\begin{figure}[H]
    \centering
    \includegraphics[width=\columnwidth]{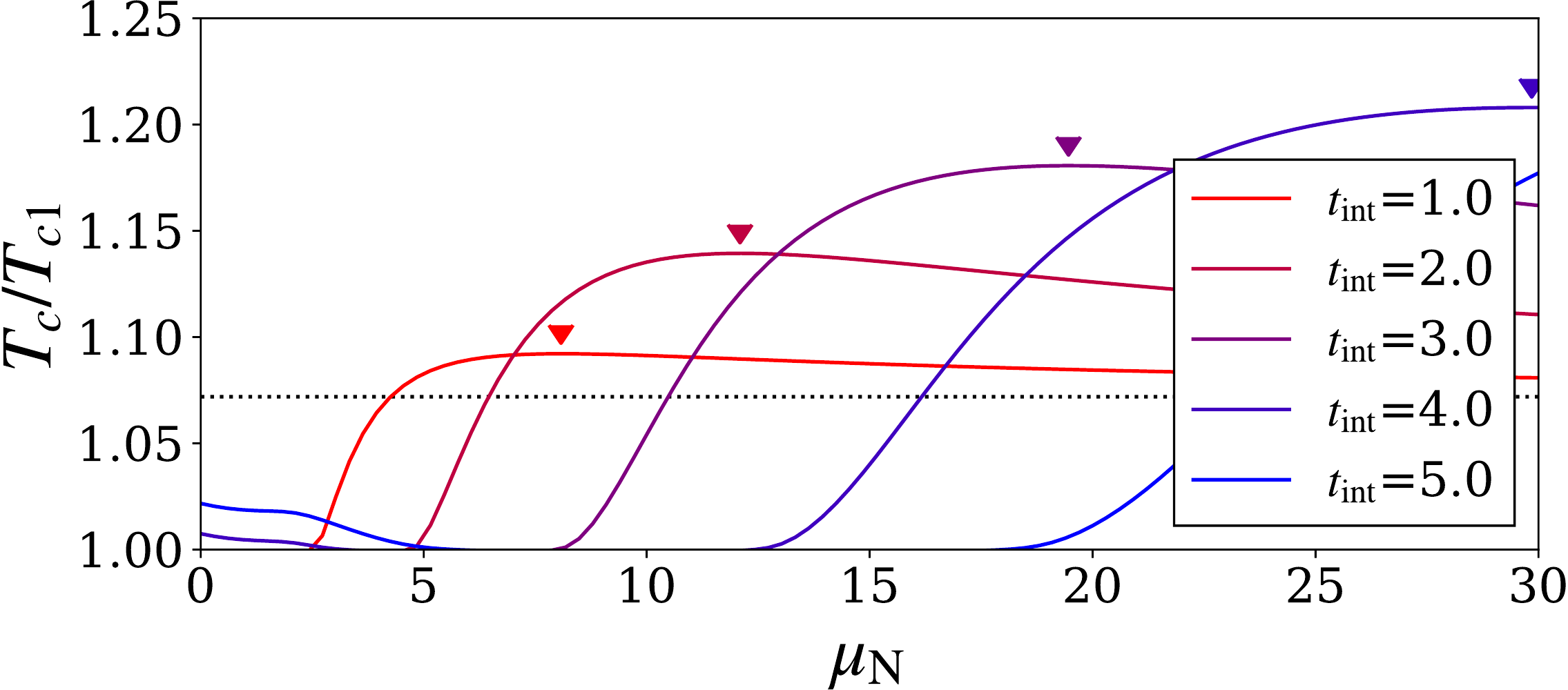}
    \includegraphics[width=\columnwidth]{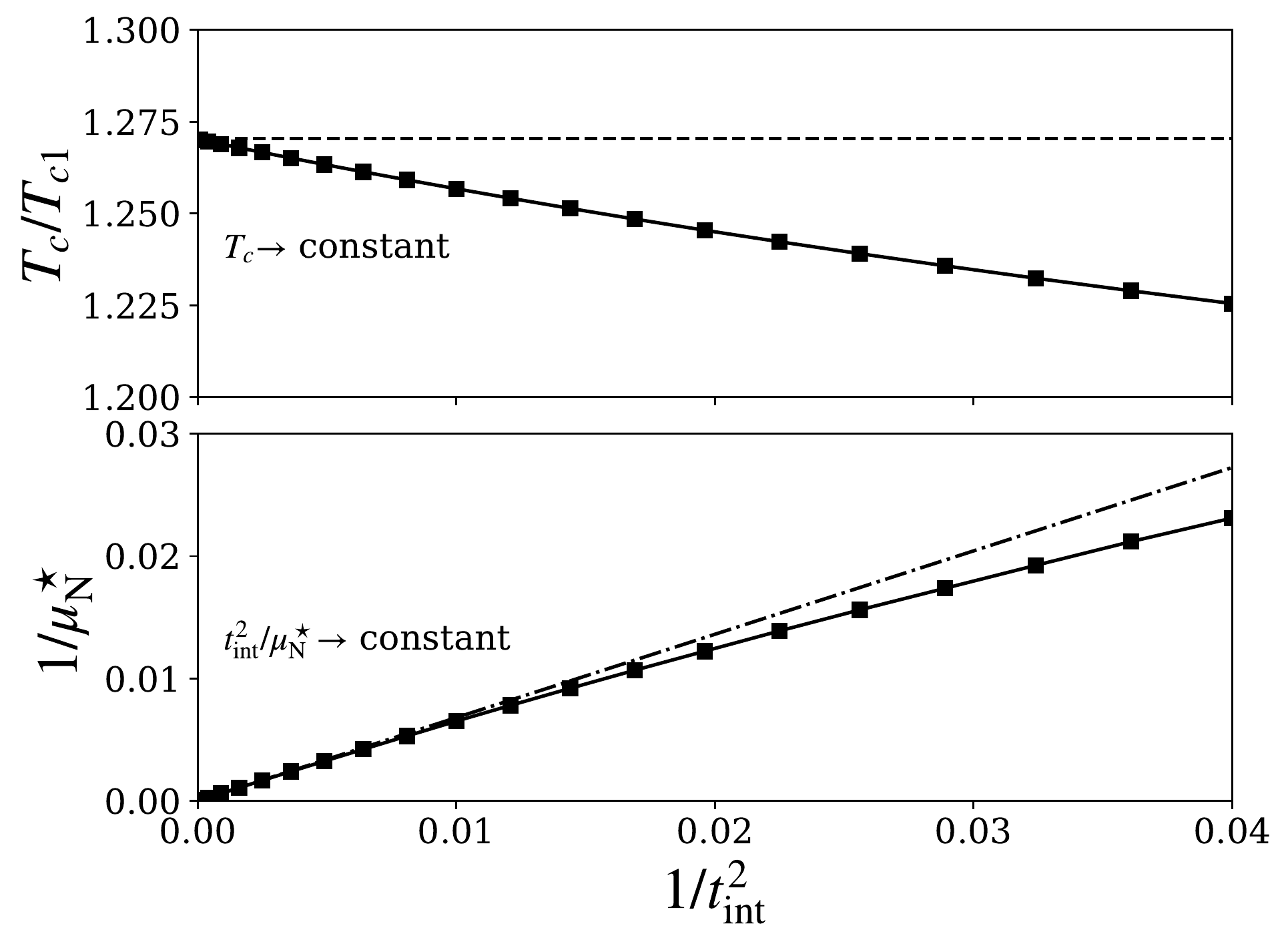}
    \caption{
    Upper panel: Interface critical temperature $T_c$ as a function of the normal on-site potential $\mu_\normal$ for various interface hopping parameters $t_\inter$. The triangles mark the optimal on-site potential $\mu_\normal^\star$ where the critical temperature is maximal. Both the optimal on-site potential $\mu_\normal^\star$ and the associated critical temperature increases as $t_\inter$ increases. For comparison, the dotted line corresponds to the hard-wall critical temperature $T_{c2}$. Middle and lower panel: Maximal critical temperature and optimal on-site potential for large $t_\inter$. The critical temperature approaches its upper bound (indicated by dashed line), while $\mu_\normal ^\star $ diverges as $t_\inter^2$ (see the dash-dotted line). Remaining parameters are set to $V=2$, $t_\SC = t_\normal = 1$, and $\mu_\SC = 1$.}
    \label{fig: example optimal mu and extrapolation}
\end{figure}

\section{Extrapolating maximal critical temperature} \label{appendix: extrapolation}

We want to determine the largest possible interface critical temperature for each on-site potential $\mu_\SC$ in the superconductor. To be specific, we want to maximize the critical temperature with respect to the remaining parameters: $t_\inter$ and $\mu_\normal$ (for simplicity we begin with fixing the hopping parameter in the normal material $t_\normal=1$). For each considered $t_\inter$, we span over $\mu_\normal$ to find the maximal critical temperature. An example of such spans is shown in \mbox{\figref{fig: example optimal mu and extrapolation}}, where $\mu_\SC=1$. For each interface hopping $t_\inter$, we locate the on-site potential $\mu_\normal ^\star$ that maximizes the critical temperature. We observe that as $t_\inter$ increases, so does $\mu_\normal ^\star$, along with the corresponding critical temperature. To find an upper bound on the possible interface critical temperature in this model, we can extrapolate the behavior for very large $t_\inter$, also shown in \mbox{\figref{fig: example optimal mu and extrapolation}} for this particular value of $\mu_\SC$. We see that the critical temperature approaches some constant for large $t_\inter$, while the optimal on-site potential $\mu_\normal^\star$ scales as $t_\inter ^2$. The limiting critical temperature and the optimal ratio $t_\inter^2 / \mu_\normal^\star$ are  independent of the hopping parameter $t_\normal$ in the normal material.

\bibliography{references.bib}

\begin{thebibliography}{23}%
\makeatletter
\providecommand \@ifxundefined [1]{%
 \@ifx{#1\undefined}
}%
\providecommand \@ifnum [1]{%
 \ifnum #1\expandafter \@firstoftwo
 \else \expandafter \@secondoftwo
 \fi
}%
\providecommand \@ifx [1]{%
 \ifx #1\expandafter \@firstoftwo
 \else \expandafter \@secondoftwo
 \fi
}%
\providecommand \natexlab [1]{#1}%
\providecommand \enquote  [1]{``#1''}%
\providecommand \bibnamefont  [1]{#1}%
\providecommand \bibfnamefont [1]{#1}%
\providecommand \citenamefont [1]{#1}%
\providecommand \href@noop [0]{\@secondoftwo}%
\providecommand \href [0]{\begingroup \@sanitize@url \@href}%
\providecommand \@href[1]{\@@startlink{#1}\@@href}%
\providecommand \@@href[1]{\endgroup#1\@@endlink}%
\providecommand \@sanitize@url [0]{\catcode `\\12\catcode `\$12\catcode
  `\&12\catcode `\#12\catcode `\^12\catcode `\_12\catcode `\%12\relax}%
\providecommand \@@startlink[1]{}%
\providecommand \@@endlink[0]{}%
\providecommand \url  [0]{\begingroup\@sanitize@url \@url }%
\providecommand \@url [1]{\endgroup\@href {#1}{\urlprefix }}%
\providecommand \urlprefix  [0]{URL }%
\providecommand \Eprint [0]{\href }%
\providecommand \doibase [0]{http://dx.doi.org/}%
\providecommand \selectlanguage [0]{\@gobble}%
\providecommand \bibinfo  [0]{\@secondoftwo}%
\providecommand \bibfield  [0]{\@secondoftwo}%
\providecommand \translation [1]{[#1]}%
\providecommand \BibitemOpen [0]{}%
\providecommand \bibitemStop [0]{}%
\providecommand \bibitemNoStop [0]{.\EOS\space}%
\providecommand \EOS [0]{\spacefactor3000\relax}%
\providecommand \BibitemShut  [1]{\csname bibitem#1\endcsname}%
\let\auto@bib@innerbib\@empty
\bibitem [{\citenamefont {De~Gennes}(1999)}]{book_de_Gennes}%
  \BibitemOpen
  \bibfield  {author} {\bibinfo {author} {\bibfnamefont {Pierre~Gilles}\
  \bibnamefont {De~Gennes}},\ }\href@noop {} {\emph {\bibinfo {title}
  {{Superconductivity of Metals and Alloys}}}},\ Advanced book classics\
  (\bibinfo  {publisher} {CRC Press},\ \bibinfo {year} {1999})\BibitemShut
  {NoStop}%
\bibitem [{\citenamefont {DE~GENNES}(1964)}]{deGennes_Boundary}%
  \BibitemOpen
  \bibfield  {author} {\bibinfo {author} {\bibfnamefont {P.~G.}\ \bibnamefont
  {DE~GENNES}},\ }\bibfield  {title} {\enquote {\bibinfo {title} {Boundary
  effects in superconductors},}\ }\href {\doibase 10.1103/RevModPhys.36.225}
  {\bibfield  {journal} {\bibinfo  {journal} {Rev. Mod. Phys.}\ }\textbf
  {\bibinfo {volume} {36}},\ \bibinfo {pages} {225--237} (\bibinfo {year}
  {1964})}\BibitemShut {NoStop}%
\bibitem [{\citenamefont {Caroli}\ \emph {et~al.}(1963)\citenamefont {Caroli},
  \citenamefont {De~Gennes},\ and\ \citenamefont {Matricon}}]{CdGM_Coherence}%
  \BibitemOpen
  \bibfield  {author} {\bibinfo {author} {\bibfnamefont {C}~\bibnamefont
  {Caroli}}, \bibinfo {author} {\bibfnamefont {PG}~\bibnamefont {De~Gennes}}, \
  and\ \bibinfo {author} {\bibfnamefont {J}~\bibnamefont {Matricon}},\
  }\bibfield  {title} {\enquote {\bibinfo {title} {Coherence length and
  penetration depth of dirty superconductors},}\ }\href@noop {} {\bibfield
  {journal} {\bibinfo  {journal} {Physik der kondensierten Materie}\ }\textbf
  {\bibinfo {volume} {1}},\ \bibinfo {pages} {176--190} (\bibinfo {year}
  {1963})}\BibitemShut {NoStop}%
\bibitem [{\citenamefont {{Caroli, C.}}\ \emph {et~al.}(1962)\citenamefont
  {{Caroli, C.}}, \citenamefont {{De Gennes, P.G.}},\ and\ \citenamefont
  {{Matricon, J.}}}]{CdGM_french}%
  \BibitemOpen
  \bibfield  {author} {\bibinfo {author} {\bibnamefont {{Caroli, C.}}},
  \bibinfo {author} {\bibnamefont {{De Gennes, P.G.}}}, \ and\ \bibinfo
  {author} {\bibnamefont {{Matricon, J.}}},\ }\bibfield  {title} {\enquote
  {\bibinfo {title} {Sur certaines propri\'et\'es des alliages supraconducteurs
  non magn\'etiques},}\ }\href {\doibase 10.1051/jphysrad:019620023010070700}
  {\bibfield  {journal} {\bibinfo  {journal} {J. Phys. Radium}\ }\textbf
  {\bibinfo {volume} {23}},\ \bibinfo {pages} {707--716} (\bibinfo {year}
  {1962})}\BibitemShut {NoStop}%
\bibitem [{\citenamefont {Abrikosov}(1965)}]{abrikosov1965concerning}%
  \BibitemOpen
  \bibfield  {author} {\bibinfo {author} {\bibfnamefont {AA}~\bibnamefont
  {Abrikosov}},\ }\bibfield  {title} {\enquote {\bibinfo {title} {Concerning
  surface superconductivity in strong magnetic fields},}\ }\href@noop {}
  {\bibfield  {journal} {\bibinfo  {journal} {Sov. Phys. JETP}\ }\textbf
  {\bibinfo {volume} {20}},\ \bibinfo {pages} {480} (\bibinfo {year}
  {1965})}\BibitemShut {NoStop}%
\bibitem [{\citenamefont {Andryushin}\ \emph {et~al.}(1993)\citenamefont
  {Andryushin}, \citenamefont {Ginzburg},\ and\ \citenamefont
  {Silin}}]{andryushin1993boundary}%
  \BibitemOpen
  \bibfield  {author} {\bibinfo {author} {\bibfnamefont {Evgenii~A}\
  \bibnamefont {Andryushin}}, \bibinfo {author} {\bibfnamefont {Vitalii~L}\
  \bibnamefont {Ginzburg}}, \ and\ \bibinfo {author} {\bibfnamefont
  {Andrei~Pavlovich}\ \bibnamefont {Silin}},\ }\bibfield  {title} {\enquote
  {\bibinfo {title} {Boundary conditions in the macroscopic theory of
  superconductivity},}\ }\href@noop {} {\bibfield  {journal} {\bibinfo
  {journal} {Physics-Uspekhi}\ }\textbf {\bibinfo {volume} {36}},\ \bibinfo
  {pages} {854} (\bibinfo {year} {1993})}\BibitemShut {NoStop}%
\bibitem [{\citenamefont {Shapoval}(1985)}]{shapoval1985boundary}%
  \BibitemOpen
  \bibfield  {author} {\bibinfo {author} {\bibfnamefont {EA}~\bibnamefont
  {Shapoval}},\ }\bibfield  {title} {\enquote {\bibinfo {title} {Boundary
  conditions on the ginzburg-landau equations for anisotropic
  superconductors},}\ }\href@noop {} {\bibfield  {journal} {\bibinfo  {journal}
  {Zh. Eksp. Teor. Fiz}\ }\textbf {\bibinfo {volume} {88}},\ \bibinfo {pages}
  {1073--1078} (\bibinfo {year} {1985})}\BibitemShut {NoStop}%
\bibitem [{\citenamefont {Samoilenka}\ and\ \citenamefont
  {Babaev}(2020)}]{samoilenka2020boundary}%
  \BibitemOpen
  \bibfield  {author} {\bibinfo {author} {\bibfnamefont {Albert}\ \bibnamefont
  {Samoilenka}}\ and\ \bibinfo {author} {\bibfnamefont {Egor}\ \bibnamefont
  {Babaev}},\ }\bibfield  {title} {\enquote {\bibinfo {title} {Boundary states
  with elevated critical temperatures in bardeen-cooper-schrieffer
  superconductors},}\ }\href {\doibase 10.1103/PhysRevB.101.134512} {\bibfield
  {journal} {\bibinfo  {journal} {Phys. Rev. B}\ }\textbf {\bibinfo {volume}
  {101}},\ \bibinfo {pages} {134512} (\bibinfo {year} {2020})}\BibitemShut
  {NoStop}%
\bibitem [{\citenamefont {Samoilenka}\ \emph {et~al.}(2020)\citenamefont
  {Samoilenka}, \citenamefont {Barkman}, \citenamefont {Benfenati},\ and\
  \citenamefont {Babaev}}]{samoilenka2020pair}%
  \BibitemOpen
  \bibfield  {author} {\bibinfo {author} {\bibfnamefont {Albert}\ \bibnamefont
  {Samoilenka}}, \bibinfo {author} {\bibfnamefont {Mats}\ \bibnamefont
  {Barkman}}, \bibinfo {author} {\bibfnamefont {Andrea}\ \bibnamefont
  {Benfenati}}, \ and\ \bibinfo {author} {\bibfnamefont {Egor}\ \bibnamefont
  {Babaev}},\ }\bibfield  {title} {\enquote {\bibinfo {title}
  {Pair-density-wave superconductivity of faces, edges, and vertices in systems
  with imbalanced fermions},}\ }\href {\doibase 10.1103/PhysRevB.101.054506}
  {\bibfield  {journal} {\bibinfo  {journal} {Phys. Rev. B}\ }\textbf {\bibinfo
  {volume} {101}},\ \bibinfo {pages} {054506} (\bibinfo {year}
  {2020})}\BibitemShut {NoStop}%
\bibitem [{\citenamefont {Samoilenka}\ and\ \citenamefont
  {Babaev}(2021)}]{samoilenka2021microscopic}%
  \BibitemOpen
  \bibfield  {author} {\bibinfo {author} {\bibfnamefont {Albert}\ \bibnamefont
  {Samoilenka}}\ and\ \bibinfo {author} {\bibfnamefont {Egor}\ \bibnamefont
  {Babaev}},\ }\bibfield  {title} {\enquote {\bibinfo {title} {Microscopic
  derivation of superconductor-insulator boundary conditions for
  ginzburg-landau theory revisited: Enhanced superconductivity at boundaries
  with and without magnetic field},}\ }\href {\doibase
  10.1103/PhysRevB.103.224516} {\bibfield  {journal} {\bibinfo  {journal}
  {Phys. Rev. B}\ }\textbf {\bibinfo {volume} {103}},\ \bibinfo {pages}
  {224516} (\bibinfo {year} {2021})}\BibitemShut {NoStop}%
\bibitem [{\citenamefont {Hainzl}\ \emph {et~al.}(2022)\citenamefont {Hainzl},
  \citenamefont {Roos},\ and\ \citenamefont {Seiringer}}]{hainzl2022boundary}%
  \BibitemOpen
  \bibfield  {author} {\bibinfo {author} {\bibfnamefont {Christian}\
  \bibnamefont {Hainzl}}, \bibinfo {author} {\bibfnamefont {Barbara}\
  \bibnamefont {Roos}}, \ and\ \bibinfo {author} {\bibfnamefont {Robert}\
  \bibnamefont {Seiringer}},\ }\href@noop {} {\enquote {\bibinfo {title}
  {Boundary superconductivity in the bcs model},}\ } (\bibinfo {year} {2022}),\
  \Eprint {http://arxiv.org/abs/2201.08090} {arXiv:2201.08090 [math-ph]}
  \BibitemShut {NoStop}%
\bibitem [{\citenamefont {Fink}\ and\ \citenamefont
  {Joiner}(1969)}]{fink1969surface}%
  \BibitemOpen
  \bibfield  {author} {\bibinfo {author} {\bibfnamefont {H.~J.}\ \bibnamefont
  {Fink}}\ and\ \bibinfo {author} {\bibfnamefont {W.~C.~H.}\ \bibnamefont
  {Joiner}},\ }\bibfield  {title} {\enquote {\bibinfo {title} {Surface
  nucleation and boundary conditions in superconductors},}\ }\href {\doibase
  10.1103/PhysRevLett.23.120} {\bibfield  {journal} {\bibinfo  {journal} {Phys.
  Rev. Lett.}\ }\textbf {\bibinfo {volume} {23}},\ \bibinfo {pages} {120--123}
  (\bibinfo {year} {1969})}\BibitemShut {NoStop}%
\bibitem [{\citenamefont {Lortz}\ \emph {et~al.}(2006)\citenamefont {Lortz},
  \citenamefont {Tomita}, \citenamefont {Wang}, \citenamefont {Junod},
  \citenamefont {Schilling}, \citenamefont {Masui},\ and\ \citenamefont
  {Tajima}}]{lortz2006origin}%
  \BibitemOpen
  \bibfield  {author} {\bibinfo {author} {\bibfnamefont {R.}~\bibnamefont
  {Lortz}}, \bibinfo {author} {\bibfnamefont {T.}~\bibnamefont {Tomita}},
  \bibinfo {author} {\bibfnamefont {Y.}~\bibnamefont {Wang}}, \bibinfo {author}
  {\bibfnamefont {A.}~\bibnamefont {Junod}}, \bibinfo {author} {\bibfnamefont
  {J.S.}\ \bibnamefont {Schilling}}, \bibinfo {author} {\bibfnamefont
  {T.}~\bibnamefont {Masui}}, \ and\ \bibinfo {author} {\bibfnamefont
  {S.}~\bibnamefont {Tajima}},\ }\bibfield  {title} {\enquote {\bibinfo {title}
  {On the origin of the double superconducting transition in overdoped
  yba2cu3ox},}\ }\href {\doibase https://doi.org/10.1016/j.physc.2005.12.066}
  {\bibfield  {journal} {\bibinfo  {journal} {Physica C: Superconductivity}\
  }\textbf {\bibinfo {volume} {434}},\ \bibinfo {pages} {194--198} (\bibinfo
  {year} {2006})}\BibitemShut {NoStop}%
\bibitem [{\citenamefont {Janod}\ \emph {et~al.}(1993)\citenamefont {Janod},
  \citenamefont {Junod}, \citenamefont {Graf}, \citenamefont {Wang},
  \citenamefont {Triscone},\ and\ \citenamefont {Muller}}]{janod1993split}%
  \BibitemOpen
  \bibfield  {author} {\bibinfo {author} {\bibfnamefont {E.}~\bibnamefont
  {Janod}}, \bibinfo {author} {\bibfnamefont {A.}~\bibnamefont {Junod}},
  \bibinfo {author} {\bibfnamefont {T.}~\bibnamefont {Graf}}, \bibinfo {author}
  {\bibfnamefont {K.-Q.}\ \bibnamefont {Wang}}, \bibinfo {author}
  {\bibfnamefont {G.}~\bibnamefont {Triscone}}, \ and\ \bibinfo {author}
  {\bibfnamefont {J.}~\bibnamefont {Muller}},\ }\bibfield  {title} {\enquote
  {\bibinfo {title} {Split superconducting transitions in the specific heat and
  magnetic susceptibility of yba2cu3ox versus oxygen content},}\ }\href
  {\doibase https://doi.org/10.1016/0921-4534(93)90643-5} {\bibfield  {journal}
  {\bibinfo  {journal} {Physica C: Superconductivity}\ }\textbf {\bibinfo
  {volume} {216}},\ \bibinfo {pages} {129--139} (\bibinfo {year}
  {1993})}\BibitemShut {NoStop}%
\bibitem [{\citenamefont {Khlyustikov}(2011)}]{khlyustikov2011critical}%
  \BibitemOpen
  \bibfield  {author} {\bibinfo {author} {\bibfnamefont {I~N}\ \bibnamefont
  {Khlyustikov}},\ }\bibfield  {title} {\enquote {\bibinfo {title} {{Critical
  magnetic field of surface superconductivity in lead}},}\ }\href {\doibase
  10.1134/S1063776111140056} {\bibfield  {journal} {\bibinfo  {journal}
  {Journal of Experimental and Theoretical Physics}\ }\textbf {\bibinfo
  {volume} {113}},\ \bibinfo {pages} {1032--1034} (\bibinfo {year}
  {2011})}\BibitemShut {NoStop}%
\bibitem [{\citenamefont {Khlyustikov}(2016)}]{khlyustikov2016surface}%
  \BibitemOpen
  \bibfield  {author} {\bibinfo {author} {\bibfnamefont {I~N}\ \bibnamefont
  {Khlyustikov}},\ }\bibfield  {title} {\enquote {\bibinfo {title} {{Surface
  superconductivity in lead}},}\ }\href {\doibase 10.1134/S1063776116020047}
  {\bibfield  {journal} {\bibinfo  {journal} {Journal of Experimental and
  Theoretical Physics}\ }\textbf {\bibinfo {volume} {122}},\ \bibinfo {pages}
  {328--330} (\bibinfo {year} {2016})}\BibitemShut {NoStop}%
\bibitem [{\citenamefont {Kozhevnikov}\ \emph {et~al.}(2007)\citenamefont
  {Kozhevnikov}, \citenamefont {Bael}, \citenamefont {Sahoo}, \citenamefont
  {Temst}, \citenamefont {Haesendonck}, \citenamefont {Vantomme},\ and\
  \citenamefont {Indekeu}}]{kozhevnikov2007observation}%
  \BibitemOpen
  \bibfield  {author} {\bibinfo {author} {\bibfnamefont {V~F}\ \bibnamefont
  {Kozhevnikov}}, \bibinfo {author} {\bibfnamefont {M~J~Van}\ \bibnamefont
  {Bael}}, \bibinfo {author} {\bibfnamefont {P~K}\ \bibnamefont {Sahoo}},
  \bibinfo {author} {\bibfnamefont {K}~\bibnamefont {Temst}}, \bibinfo {author}
  {\bibfnamefont {C~Van}\ \bibnamefont {Haesendonck}}, \bibinfo {author}
  {\bibfnamefont {A}~\bibnamefont {Vantomme}}, \ and\ \bibinfo {author}
  {\bibfnamefont {J~O}\ \bibnamefont {Indekeu}},\ }\bibfield  {title} {\enquote
  {\bibinfo {title} {Observation of wetting-like phase transitions in a
  surface-enhanced type-i superconductor},}\ }\href {\doibase
  10.1088/1367-2630/9/3/075} {\bibfield  {journal} {\bibinfo  {journal} {New
  Journal of Physics}\ }\textbf {\bibinfo {volume} {9}},\ \bibinfo {pages}
  {75--75} (\bibinfo {year} {2007})}\BibitemShut {NoStop}%
\bibitem [{\citenamefont {Khlyustikov}(2021)}]{khlyustikov2021surface}%
  \BibitemOpen
  \bibfield  {author} {\bibinfo {author} {\bibfnamefont {I~N}\ \bibnamefont
  {Khlyustikov}},\ }\bibfield  {title} {\enquote {\bibinfo {title} {{Surface
  Superconductivity of Vanadium}},}\ }\href {\doibase
  10.1134/S1063776121030043} {\bibfield  {journal} {\bibinfo  {journal}
  {Journal of Experimental and Theoretical Physics}\ }\textbf {\bibinfo
  {volume} {132}},\ \bibinfo {pages} {453--456} (\bibinfo {year}
  {2021})}\BibitemShut {NoStop}%
\bibitem [{\citenamefont {Mangel}\ \emph {et~al.}(2020)\citenamefont {Mangel},
  \citenamefont {Kapon}, \citenamefont {Blau}, \citenamefont {Golubkov},
  \citenamefont {Gavish},\ and\ \citenamefont
  {Keren}}]{mangel2020stiffnessometer}%
  \BibitemOpen
  \bibfield  {author} {\bibinfo {author} {\bibfnamefont {Itay}\ \bibnamefont
  {Mangel}}, \bibinfo {author} {\bibfnamefont {Itzik}\ \bibnamefont {Kapon}},
  \bibinfo {author} {\bibfnamefont {Nitzan}\ \bibnamefont {Blau}}, \bibinfo
  {author} {\bibfnamefont {Katrine}\ \bibnamefont {Golubkov}}, \bibinfo
  {author} {\bibfnamefont {Nir}\ \bibnamefont {Gavish}}, \ and\ \bibinfo
  {author} {\bibfnamefont {Amit}\ \bibnamefont {Keren}},\ }\bibfield  {title}
  {\enquote {\bibinfo {title} {Stiffnessometer: A magnetic-field-free
  superconducting stiffness meter and its application},}\ }\href {\doibase
  10.1103/PhysRevB.102.024502} {\bibfield  {journal} {\bibinfo  {journal}
  {Phys. Rev. B}\ }\textbf {\bibinfo {volume} {102}},\ \bibinfo {pages}
  {024502} (\bibinfo {year} {2020})}\BibitemShut {NoStop}%
\bibitem [{\citenamefont {Tsindlekht}\ \emph {et~al.}(2004)\citenamefont
  {Tsindlekht}, \citenamefont {Leviev}, \citenamefont {Asulin}, \citenamefont
  {Sharoni}, \citenamefont {Millo}, \citenamefont {Felner}, \citenamefont
  {Paderno}, \citenamefont {Filippov},\ and\ \citenamefont
  {Belogolovskii}}]{tsindlekht2004tunneling}%
  \BibitemOpen
  \bibfield  {author} {\bibinfo {author} {\bibfnamefont {M.~I.}\ \bibnamefont
  {Tsindlekht}}, \bibinfo {author} {\bibfnamefont {G.~I.}\ \bibnamefont
  {Leviev}}, \bibinfo {author} {\bibfnamefont {I.}~\bibnamefont {Asulin}},
  \bibinfo {author} {\bibfnamefont {A.}~\bibnamefont {Sharoni}}, \bibinfo
  {author} {\bibfnamefont {O.}~\bibnamefont {Millo}}, \bibinfo {author}
  {\bibfnamefont {I.}~\bibnamefont {Felner}}, \bibinfo {author} {\bibfnamefont
  {Yu.~B.}\ \bibnamefont {Paderno}}, \bibinfo {author} {\bibfnamefont {V.~B.}\
  \bibnamefont {Filippov}}, \ and\ \bibinfo {author} {\bibfnamefont {M.~A.}\
  \bibnamefont {Belogolovskii}},\ }\bibfield  {title} {\enquote {\bibinfo
  {title} {Tunneling and magnetic characteristics of superconducting
  ${\mathrm{zrb}}_{12}$ single crystals},}\ }\href {\doibase
  10.1103/PhysRevB.69.212508} {\bibfield  {journal} {\bibinfo  {journal} {Phys.
  Rev. B}\ }\textbf {\bibinfo {volume} {69}},\ \bibinfo {pages} {212508}
  (\bibinfo {year} {2004})}\BibitemShut {NoStop}%
\bibitem [{\citenamefont {Belogolovskii}\ \emph {et~al.}(2011)\citenamefont
  {Belogolovskii}, \citenamefont {Felner},\ and\ \citenamefont
  {Shaternik}}]{belogolovskii2010zirconium}%
  \BibitemOpen
  \bibfield  {author} {\bibinfo {author} {\bibfnamefont {Mikhail}\ \bibnamefont
  {Belogolovskii}}, \bibinfo {author} {\bibfnamefont {Israel}\ \bibnamefont
  {Felner}}, \ and\ \bibinfo {author} {\bibfnamefont {Vladimir}\ \bibnamefont
  {Shaternik}},\ }\bibfield  {title} {\enquote {\bibinfo {title} {Zirconium
  dodecaboride, a novel superconducting material with enhanced surface
  characteristics},}\ }in\ \href@noop {} {\emph {\bibinfo {booktitle} {Boron
  Rich Solids}}},\ \bibinfo {editor} {edited by\ \bibinfo {editor}
  {\bibfnamefont {Nina}\ \bibnamefont {Orlovskaya}}\ and\ \bibinfo {editor}
  {\bibfnamefont {Mykola}\ \bibnamefont {Lugovy}}}\ (\bibinfo  {publisher}
  {Springer Netherlands},\ \bibinfo {address} {Dordrecht},\ \bibinfo {year}
  {2011})\ pp.\ \bibinfo {pages} {195--206}\BibitemShut {NoStop}%
\bibitem [{\citenamefont {Khasanov}\ \emph {et~al.}(2005)\citenamefont
  {Khasanov}, \citenamefont {Di~Castro}, \citenamefont {Belogolovskii},
  \citenamefont {Paderno}, \citenamefont {Filippov}, \citenamefont
  {Br\"utsch},\ and\ \citenamefont {Keller}}]{khasanov2005anomalous}%
  \BibitemOpen
  \bibfield  {author} {\bibinfo {author} {\bibfnamefont {R.}~\bibnamefont
  {Khasanov}}, \bibinfo {author} {\bibfnamefont {D.}~\bibnamefont {Di~Castro}},
  \bibinfo {author} {\bibfnamefont {M.}~\bibnamefont {Belogolovskii}}, \bibinfo
  {author} {\bibfnamefont {Yu.}\ \bibnamefont {Paderno}}, \bibinfo {author}
  {\bibfnamefont {V.}~\bibnamefont {Filippov}}, \bibinfo {author}
  {\bibfnamefont {R.}~\bibnamefont {Br\"utsch}}, \ and\ \bibinfo {author}
  {\bibfnamefont {H.}~\bibnamefont {Keller}},\ }\bibfield  {title} {\enquote
  {\bibinfo {title} {Anomalous electron-phonon coupling probed on the surface
  of superconductor $\mathrm{Zr}{\mathrm{b}}_{12}$},}\ }\href {\doibase
  10.1103/PhysRevB.72.224509} {\bibfield  {journal} {\bibinfo  {journal} {Phys.
  Rev. B}\ }\textbf {\bibinfo {volume} {72}},\ \bibinfo {pages} {224509}
  (\bibinfo {year} {2005})}\BibitemShut {NoStop}%
\bibitem [{\citenamefont {Ginzburg}(1964)}]{ginzburg1964surface}%
  \BibitemOpen
  \bibfield  {author} {\bibinfo {author} {\bibfnamefont {V.L.}\ \bibnamefont
  {Ginzburg}},\ }\bibfield  {title} {\enquote {\bibinfo {title} {On surface
  superconductivity},}\ }\href {\doibase
  https://doi.org/10.1016/0031-9163(64)90672-9} {\bibfield  {journal} {\bibinfo
   {journal} {Physics Letters}\ }\textbf {\bibinfo {volume} {13}},\ \bibinfo
  {pages} {101--102} (\bibinfo {year} {1964})}\BibitemShut {NoStop}%
\end{thebibliography}%

\end{document}